# New Interpretation for error propagation of data-driven Reynolds stress closures via global stability analysis

Xianglin Shan[1,2], Wenbo Cao[1,2], Weiwei Zhang[1,2,*]

[1]School of Aeronautics, Northwestern Polytechnical University, Xi'an 710072, China

[2] International Joint Institute of Artificial Intelligence on Fluid Mechanics, Northwestern Polytechnical University, Xi'an, 710072, China

In light of the challenges surrounding convergence and error propagation encountered in Reynolds-averaged Navier-Stokes (RANS) equations with data-driven Reynolds stress closures, researchers commonly attribute these issues to ill-conditioning through conditional number analysis. This paper delves into an additional factor, numerical instability, contributing to these challenges. We conduct global stability analysis for the RANS equations, closed by the Reynolds stress of direct numerical simulation (DNS), with the time-averaged solution of DNS as the base flow. Our findings reveal that, for turbulent channel flow at high Reynolds numbers, significant ill-conditioning exists, yet the system remains stable. Conversely, for separated flow over periodic hills, notable ill-conditioning is absent, but unstable eigenvalues are present, indicating that error propagation arises from the mechanism of numerical instability. Furthermore, the effectiveness of the decomposition method employing eddy viscosity is compared, results show that the spatial distribution and amplitude of eddy viscosity influences the numerical stability.

## I. INTRODUCTION

Turbulence is the most complicated kind of the fluid motion and a problem of the century in classical physics [1]. Among the numerical simulation methods of turbulence, direct numerical simulation (DNS) and large eddy simulation (LES) are costly, while Reynolds-averaged Navier–Stokes (RANS) equations with turbulence models will continue to remain the critical approaches in industrial simulations for the foreseeable future [2,3]. The turbulence model theory is a semi-empirical theory, which absorbs the results of turbulence statistical theory, turbulence experiments and direct numerical simulations, and completes Reynolds stress closures with the aid of turbulence data and physical analogy. The turbulent model theory mainly includes eddy viscosity model (EVM) and Reynolds stress transport model (RSTM). Eddy viscosity model, e.g., Spalart–Allmaras (SA) [4], Shear Stress Transport (SST) model [5], has been widely used in the industrial community, but cannot provide satisfactory predictive accuracy in many complex flows with streamline curvature, strong adverse pressure gradients, and separation[6]. RSTM is more accurate than EVM in complex flows, but is poorly converged and robust, significantly restricting its industrial applications.

---

[*] Corresponding author: aeroelastic@nwpu.edu.cn

Recently, data-driven methods have been increasingly applied to turbulence modeling with the help of machine learning (ML) and high-fidelity data [7,8]. Researchers has proposed many data-driven eddy viscosity models [9-12], Reynolds stress models [13-16], and Reynolds force models [17,18]. For data-driven Reynolds stress models, a bottleneck is the **error propagation** problem, i.e., substituting the Reynolds stress from DNS into the RANS equations can lead to large error of velocity. This phenomenon was found by Thompson et al. [19]. They studied turbulent channel flows and presented the error propagation equation,

$$E_u(y) = Re_\tau \int_0^y E_R(y')dy' \tag{1}$$

where $E_R(y')$ is the error of Reynolds stress, $E_u(y)$ is the error of velocity, and $Re_\tau$ is the friction Reynolds number. Using Eq.(1) and DNS data of Lee & Moser [20], the error propagation factor of turbulent channel flow with different Reynolds number (Re) can be calculated as shown in Fig.1. It can be found that the error propagation factor is approximatively positively correlated with Re.

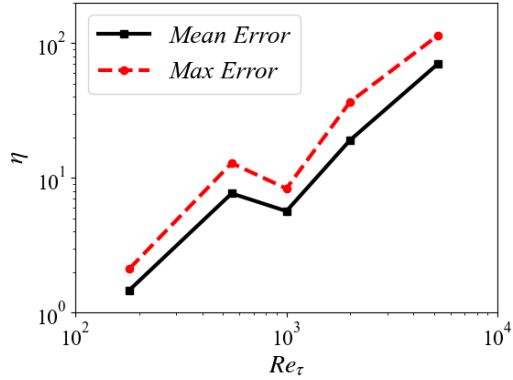

Fig.1 Error propagation factor of turbulent channel flow with different Re. For mean error, $\eta$ is calculated by $\eta = \dfrac{\int_0^h E_u(y)dy}{\int_0^h E_R(y)dy}$, where $h$ denotes the half channel width. For maximal error, $\eta$ is calculated by $\eta = \dfrac{\max\limits_{0 \leq y \leq h}(E_u(y))}{\max\limits_{0 \leq y \leq h}(E_R(y))}$.

Therefore, to overcome this bottleneck, the reason of error propagation and coupling calculation methods of decreasing the error are two key problems. Wu et al. [14,21] verified error propagation phenomenon via numerical simulation of three cases, i.e., turbulent channel flow, separated flow over periodic hills, and flow in a square duct, and regarded the reason as **ill-conditioning**. They proposed a local condition number function for evaluation of the conditioning of the RANS equations. To reduce the error, they proposed a new coupling method between RANS equations and Reynolds stress, i.e., decomposing the Reynolds stress into linear and nonlinear parts

$$\tau = 2v_t S + \tau^\perp \tag{2}$$

where $v_t$ is the eddy viscosity, and $\tau$ is the Reynolds stress and $S$ is the strain rate tensor of DNS data. $2v_t S$ is the linear part, and $\tau^\perp$ is the nonlinear part. Their results show that the condition number is significantly reduced by this method, and the error of velocity is reduced. Brener et al. [17] deduced the error propagation equation of Wu's method for turbulent channel flow and interpreted the advantageous effect of positive eddy viscosity. They also thought the possible ill-conditioning results in error propagation. Guo et al. [22,23] studied the decomposition method of Reynolds stress with eddy viscosity from different turbulence models and achieved a new decomposition method without the information of the DNS mean velocity fields in advance.

Prior investigations have attained a profound comprehension for the reason of error propagation, i.e., ill-conditioning, and techniques for decreasing error propagation, i.e., decomposition method of Reynolds stress with eddy viscosity. However, besides ill-conditioning, other mechanisms may also lead to error propagation, and turbulent flows with different flow structures may correspond to different mechanisms. A different mechanism is the numerical instability. Although some studies, such as [24], have mentioned that the Reynolds stress models affects numerical instability, there are no studies to prove and analyze this in detail. Therefore, we hope to explore a deeper level of origin of error propagation, analyze whether the numerical instability exists, and study which factors affect ill-conditioning and numerical instability.

In this paper, we use global stability analysis (GSA) to study the ill-conditioning and numerical stability of RANS equations closed with Reynolds stress (denoted by RS) or decomposed Reynolds stress with eddy viscosity (denoted by RS-EV). The eigenvalues of the Jacobian matrices are analyzed on two representative turbulent flows, i.e., turbulent channel flow and separated flow over periodic hills. Results show that there are two different mechanisms, i.e., ill-conditioning and numerical instability, contributing to error propagation. For turbulent channel flow, ill-conditioning is the only reason, while for separated flow over periodic hills, numerical instability is the main reason.

The remainder of this article is organized as follows. Sec. II introduces global stability analysis method. Sec. III shows the results, and concluding remarks are given in Sec. IV.

## II. Methodology: Global stability analysis

Global stability analysis starts with the small amplitude perturbation waves and the core is how the perturbation waves develop. A typical flow instability analysis can be done by solving linearized Navier–Stokes equations with proper boundary conditions, which leads to an eigenvalue problem. GSA is generally used to analyze whether a flow is physically stable over time or in space. In this paper, GSA is employed to study the ill-conditioning and numerical stability of RANS equations with Reynolds stress closures.

The process of GSA is derived by the RANS equations after spatial discretization, and the small perturbation hypothesis and the Jacobian matrix are introduced. After spatial discretization, the unsteady RANS equations can be recast under the following form:

$$\frac{d\boldsymbol{q}}{dt} = R(\boldsymbol{q}) \tag{3}$$

where $\boldsymbol{q} \in \mathbb{R}^N$ represents the set of flow variables describing the flow at each spatial location of the grid. $N$ is the product of the number of grid cells and the number of flow variables. In this paper, turbulence is statistically stationary, so the time-averaged solution of DNS is the steady solution of the RANS equations that we desire, referred as base flow $\overline{\boldsymbol{q}} \in \mathbb{R}^N$, which is defined by the equation

$$R(\overline{\boldsymbol{q}}) = 0 \tag{4}$$

We consider a small amplitude perturbation $\varepsilon \boldsymbol{q}'$ superimposed on the base flow: $\boldsymbol{q} = \overline{\boldsymbol{q}} + \varepsilon \boldsymbol{q}'$, where $\varepsilon \ll 1$. The equation governing the perturbation is given by the linearization to the first order of the discretized Eq.(3):

$$\frac{d\boldsymbol{q}'}{dt} = \boldsymbol{A}\boldsymbol{q} \tag{5}$$

The global Jacobian matrix $\boldsymbol{A} \in \mathbb{R}^{N \times N}$ corresponds to the linearization of the discrete RANS residual $R$ around the base flow $\bar{\boldsymbol{q}}$:

$$A_{ij} = \left.\frac{\partial R_i}{\partial q_j}\right|_{q=\bar{q}} \tag{6}$$

where $R_i$ designates the *i*th component of the residual, which is an a priori function of all unknowns $q_j$. The global Jacobian matrix $\boldsymbol{A}$, a sparse matrix, involves spatial derivatives and is different for two coupling methods between RANS equations and Reynolds stress, i.e., RS and RS-EV.

The stability of a base flow is then determined by the spectrum of the matrix $\boldsymbol{A}$, and particular solutions of Eq. (5) are sought in the form of normal modes $\boldsymbol{q}' = \hat{\boldsymbol{q}} e^{\lambda t}$. Then, Eq. (5) may be recast into the following eigenvalue problem:

$$\boldsymbol{A}\hat{\boldsymbol{q}} = \lambda \hat{\boldsymbol{q}} \tag{7}$$

The real part of the eigenvalue $\lambda$ is the growth rate. If at least one of the eigenvalues $\lambda$ exhibits a positive growth rate, the base flow $\bar{\boldsymbol{q}}$ is unstable. Therefore, the condition of stability is that the real part of all the eigenvalues of Jacobian matrix $\boldsymbol{A}$ is less than 0, i.e.,

$$\lambda_{Real,max} < 0$$

On the other hand, large condition number of the global Jacobian matrix leads to ill-conditioning. Another approximate deduction is that, the closer the eigenvalues of $\boldsymbol{A}$ are to 0, the more ill-conditioned the RANS equations become.

The inherent stability of the system, both physical and numerical aspects, determines whether the system is stable or unstable thereafter. It is relatively easy to understand in physical aspect: for stable system, the flow can return to under the small amplitude perturbation, and it is opposite for unstable system.

Meanwhile, the aspect of numerical stability is similar. If we solve Eq.(3) with the base flow as the initial solution with time-marching method, numerical stability of the system determines whether the final solution approaches the base flow. Therefore, in this situation, numerical instability contributes to worsening error propagation and the final solution deviates system.

### III. Result

Two typical cases are investigated to discover and distinguish the origin of error propagation for different flow structures. Both these cases can be supported by time-averaged flow field data of DNS. One is turbulent channel flow at five different Reynolds numbers, typical wall turbulence. The other is the separated flow over periodic hills. While performing GSA, for numerical discretization, the viscous term is discretized with the second-order central scheme, and the convection term is discretized with a second-order upwind scheme, Roe scheme. The upwind scheme was used to avoid the possible numerical instability caused by the numerical scheme.

#### A. Turbulent channel flow

The fully developed turbulent channel flows are investigated using the DNS data of Lee & Moser [20]. A global stability analysis of the turbulent channel flow is performed. For RS-EV form, eddy viscosity is obtained by $v_t = \dfrac{\tau_{xy}}{\partial u/\partial y}$, where $\tau_{xy}$ is the *xy* component of Reynolds stress. The eigenvalues of the Jacobian matrix $A$ are calculated. As shown in Fig.2, the maximal real part of Jacobian matrix eigenvalues $\lambda_{Real,max}$ is less than 0 at five Reynolds numbers, indicating that RANS equations in both RS form and RS-EV form are stable. Meanwhile, as Re increases, $\lambda_{Real,max}$ tend to approach zero in both the RS form and RS-EV form, with the RS-EV form exhibiting a greater deviation from zero compared to the RS form. **Ill-conditioning** phenomenon can apparently be observed for RANS equations in RS form for high Re, which corresponds to the phenomenon of error propagation in Fig.1, and decomposition method of Reynolds stress greatly alleviates the degree of ill-conditioning, which is in agreement with Wu et al.[21].

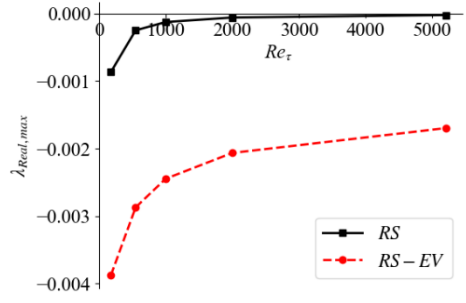

Fig.2 Maximal real part of Jacobian matrix eigenvalues for the turbulent channel flow at different Re.

## B. Separated flow over periodic hills

Separated flow over periodic hills are investigated to study the reason of error propagation for separated flows. DNS data is performed by Xiao et al.[25], with $Re_b$=5600, where $Re_b$ is defined by the bulk velocity $U_b$ at the inlet and hill height $H$. For RS-EV form, eddy viscosity is obtained by SA model and SST model. The eddy viscosity field is shown in Fig.5, revealing that the distribution of eddy viscosity vary greatly among different turbulence models. RS-EV with SA eddy viscosity is marked as RS-EV SA, and RS-EV with SST eddy viscosity is marked as RS-EV SST.

The result of global stability analysis is shown in Table 1 The real part of Jacobian matrix eigenvalues is positive for RS, while negative for both RS-EV SA and RS-EV SST. An important discovery is that the system of RANS equations in RS form is **unstable**. While the system of RANS equations in RS-EV form with both SA eddy viscosity and SST eddy viscosity is stable, and SST eddy viscosity provides more stable eigenvalue. Meanwhile, no eigenvalues tend to approach zero, ill-conditioning is not apparently detected, comparing with turbulent channel flow. Because of the existing of unstable eigenmode, iterative system cannot converge to the DNS time-averaged solution for the system of RANS equations in RS form, as reported in [17,21]. Therefore , **instability** can be an new interpretation for error propagation of the system of RANS equations is different from the interpretation of Wu et al.[21] and Brener et al.[17]. Meanwhile, the different spatial distribution of eddy viscosity from SA and SST also influences the numerical stability.

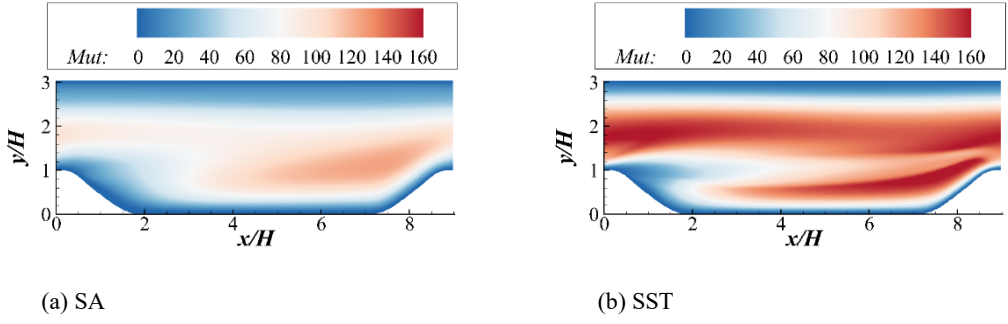

(a) SA  (b) SST

Fig.3 Eddy viscosity field from different turbulence models for separated flow over periodic hills

Table 1 Maximal real part of Jacobian matrix eigenvalues with different decomposition methods

| Decomposition method | $\lambda_{Real,max}$ |
| --- | --- |
| RS | 0.072 |
| RS-EV SA | −0.258 |
| RS-EV SST | −0.877 |

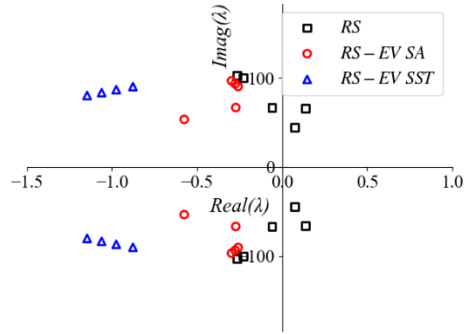

Fig. 4 Several Jacobian matrix eigenvalues with maximal real part with different decomposition methods

To study effect of the amplitude of eddy viscosity on the numerical stability of the system of RANS equations, we apply the decomposition

$$\boldsymbol{\tau} = 2\eta v_t \boldsymbol{S} + \boldsymbol{\tau}^{\perp} \qquad (8)$$

with an adjustable parameter $\eta$. $\eta$ is set from $10^{-3}$ to $10^{0}$, and can decrease the amplitude of eddy viscosity. The stability of RANS system with different $\eta$ is shown in Fig.4. As $\eta$ increase, the eddy viscosity part of Eq.(8) increases. $\eta \to 0$ means the decomposition system tend to the system in RS form, while $\eta \to 1$ means the decomposition system tend to the system in RS-EV form. When $\eta$ is over 1, the system can be more stable. Therefore, the amplitude of eddy viscosity determines the numerical stability of RANS system.

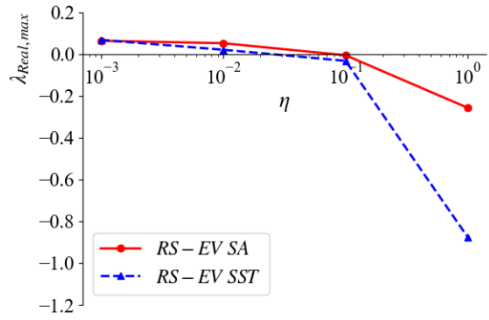

Fig.5 Maximal real part of Jacobian matrix eigenvalues for different $\eta$ for separated flow over periodic hills

## VI. CONCLUSION

Applying global stability analysis, this paper discovers a new explanation for the error propagation of RANS equations with Reynolds stress closures. From the eigenvalues of the global Jacobian matrices, we explore the different reasons for the error propagation of turbulent channel flow and separated flow over periodic hills.

The error propagation of turbulent channel flow is mainly caused by ill-conditioning, while the error propagation of separated flow over periodic hills is mainly caused by numerical instability. The different flow structures of these two types of flows affect the global Jacobian matrix of the system, which affects the ill-conditioning and numerical stability. The key factors affect ill-conditioning and numerical stability are Reynolds number and decomposition method. With Re increasing, the system tends to more ill-conditioned. For the decomposition method of Reynolds stress, the spatial distribution of eddy viscosity has an impact on the stability of the system, and larger amplitude of eddy viscosity corresponds to a more stable system.

## ACKNOWLEDGMENTS

This paper was supported by National Natural Science Foundation of China (No. 92152301).

## REFERENCE

[1] P. Moin and J. Kim, Tackling turbulence with supercomputers, Sci. Am. **276**, 62 (1997).

[2] P. A. Durbin, Some recent developments in turbulence closure modeling, Annu. Rev. Fluid Mech. **50**, 77 (2018).

[3] J. P. Slotnick, A. Khodadoust, J. Alonso, D. Darmofal, W. Gropp, E. Lurie, and D. J. Mavriplis, Cfd vision

2030 study: A path to revolutionary computational aerosciences, 2014.

[4] P. Spalart and S. Allmaras, A one-equation turbulence model for aerodynamic flows in *30th aerospace sciences meeting and exhibit*1992), p. 439.

[5] F. R. Menter, 2-equation eddy-viscosity turbulence models for engineering applications, AIAA J. **32**, 1598 (1994).

[6] T. J. Craft, B. E. Launder, and K. Suga, Development and application of a cubic eddy-viscosity model of turbulence, Int. J. Heat Fluid Flow **17**, 108 (1996).

[7] K. Duraisamy, G. Iaccarino, and H. Xiao, Turbulence modeling in the age of data, Annu. Rev. Fluid Mech. **51**, 357 (2019).

[8] K. Duraisamy, Perspectives on machine learning-augmented reynolds-averaged and large eddy simulation models of turbulence, Phys. Rev. Fluids **6**, 050504 (2021).

[9] A. P. Singh, S. Medida, and K. Duraisamy, Machine-learning-augmented predictive modeling of turbulent separated flows over airfoils, AIAA J. **55**, 2215 (2017).

[10] A. P. Singh and K. Duraisamy, Using field inversion to quantify functional errors in turbulence closures, Phys. Fluids **28**, 045110 (2016).

[11] W. Liu, J. Fang, S. Rolfo, C. Moulinec, and D. R. Emerson, On the improvement of the extrapolation capability of an iterative machine-learning based rans framework, Comput. Fluids **256**, 105864 (2023).

[12] X. Shan, Y. Liu, W. Cao, X. Sun, and W. Zhang, Turbulence modeling via data assimilation and machine learning for separated flows over airfoils, AIAA J. **61**, 3883 (2023).

[13] J.-X. Wang, J.-L. Wu, and H. Xiao, Physics-informed machine learning approach for reconstructing reynolds stress modeling discrepancies based on dns data, Phys. Rev. Fluids **2**, 034603 (2017).

[14] J.-L. Wu, H. Xiao, and E. Paterson, Physics-informed machine learning approach for augmenting turbulence models: A comprehensive framework, Phys. Rev. Fluids **3**, 074602 (2018).

[15] J. Ling, A. Kurzawski, and J. Templeton, Reynolds averaged turbulence modelling using deep neural networks with embedded invariance, J. Fluid Mech. **807**, 155 (2016).

[16] J. Weatheritt and R. Sandberg, A novel evolutionary algorithm applied to algebraic modifications of the rans stress-strain relationship, J. Comput. Phys. **325**, 22 (2016).

[17] B. P. Brener, M. A. Cruz, R. L. Thompson, and R. P. Anjos, Conditioning and accurate solutions of reynolds average navier-stokes equations with data-driven turbulence closures, J. Fluid Mech. **915**, A110 (2021).

[18] M. A. Cruz, R. L. Thompson, L. E. B. Sampaio, and R. D. A. Bacchi, The use of the reynolds force vector in a physics informed machine learning approach for predictive turbulence modeling, Comput. Fluids **192**, 104258 (2019).

[19] R. L. Thompson, L. E. B. Sampaio, F. A. V. de Braganca Alves, L. Thais, and G. Mompean, A methodology to evaluate statistical errors in dns data of plane channel flows, Comput. Fluids **130**, 1 (2016).

[20] M. Lee and R. D. Moser, Direct numerical simulation of turbulent channel flow up to, J. Fluid Mech. **774**, 395 (2015).

[21] J. Wu, H. Xiao, R. Sun, and Q. Wang, Reynolds-averaged navier–stokes equations with explicit data-driven reynolds stress closure can be ill-conditioned, J. Fluid Mech. **869**, 553 (2019).

[22] X. Guo, Z. Xia, and S. Chen, Practical framework for data-driven rans modeling with data augmentation, Acta Mechanica Sinica **37**, 1748 (2021).

[23] X. Guo, Z. Xia, and S. Chen, Computing mean fields with known reynolds stresses at steady state, Theor. Appl. Mech. Lett. **11**, 100244 (2021).

[24] M. L. A. Kaandorp and R. P. Dwight, Data-driven modelling of the reynolds stress tensor using random

forests with invariance, Comput. Fluids **202**, 104497 (2020).

[25]  H. Xiao, J.-L. Wu, S. Laizet, and L. Duan, Flows over periodic hills of parameterized geometries: A dataset for data-driven turbulence modeling from direct simulations, Comput. Fluids **200**, 104431 (2020).